\documentclass[12pt]{iopart}
\usepackage[utf8]{inputenc}
\usepackage{amssymb}
\usepackage{bm}
\usepackage{braket}
\newcommand{\text}[1]{\ensuremath{\mathrm{#1}}}
\newcommand{\sign}[0]{\ensuremath{\text{sign}}}
\usepackage{xcolor}
\usepackage{graphicx}

\begin{document}

\title{Automatic design of quantum feature maps}

\author{Sergio Altares-López {${}^1$}, Angela Ribeiro ${}^1$, Juan José García-Ripoll ${}^2$}

\address{$^1$ Centre for Automation and Robotics - CAR (CSIC-UPM), Ctra. Campo Real km. 0,200, 28500 Arganda, Spain.}
\address{$^2$ Instituto de Física Fundamental IFF-CSIC, Calle Serrano 113b, 28006 Madrid, Spain.}
\ead{juanjose.ripoll@csic.es}
\vspace{10pt}

\begin{abstract}
We propose a new technique for the automatic generation of optimal ad-hoc ansätze for classification by using quantum support vector machine (QSVM). This efficient method is based on NSGA-II multiobjective genetic algorithms which allow both maximize the accuracy and minimize the ansatz size. It is demonstrated the validity of the technique by a practical example with a non-linear dataset, interpreting the resulting circuit and its outputs. We also show other application fields of the technique that reinforce the validity of the method, and a comparison with classical classifiers in order to understand the advantages of using quantum machine learning.
\end{abstract}
\noindent{\it quantum machine learning, genetic algorithms, artificial intelligence, quantum computing, optimization, automatic quantum classifier generation \/}
\section{Introduction}

Quantum machine learning is an emerging field of research that bridges the progress in quantum computing hardware and algorithms with ideas and problems coming from artificial intelligence. The field is suffering steady and fast progress but already features a large corpus of algorithms and applications\ \cite{biamonte2017,schuld2018}. On the one hand, we find applications, such as clustering\ \cite{kerenidis2018}, quantum anomaly detection\ \cite{liu2018}, dimensionality reduction\ \cite{lloyd2014,cong2016,duan2019} or support vector machines\ \cite{rebentrost2014}, which build on the HHL algorithm for matrix inversions or related matrix-vector operations. On the other hand, we place algorithms that are ready for near-term intermediate-scale quantum (NISQ) devices, and which are typically based on parameterized quantum circuits\ \cite{benedetti2019} that implement autoencoders\ \cite{romero2017}, support vector machines and quantum classifiers\ \cite{schuld2019,havlivcek2019}, or generative adversarial networks\ \cite{dallaire-demers2018,lloyd2018}, among other applications.

Due to their simplicity and immediate experimental access, this second framework has gained considerable interest. However, a key problem of parameterized circuit is their design, both from the point of view of the structure of the circuit as well as their parameterization. The structure and design condition the expressive power of the circuit\ \cite{du2020,sim2019}, and its capacity to explore the Hilbert space and encode probability distributions more efficiently than other generative models. However, too expressive circuits can be subject to local minima and barren plateaus\ \cite{mcclean2018} that prevent reaching the optimal parameterizations. Partial solutions to these challenges include adaptive initialization strategies\ \cite{grant2019}, pruning of circuits\ \cite{sim2021}, simultaneous optimization of parameters and rotation generators\ \cite{ostaszewski2021}, or the implementation of global optimization strategies such as genetic algorithms\ \cite{li2017,lamata2018,chivilikhin2020} that optimize gates or structure.

In this work, we focus on the problem of supervised learning using quantum feature maps\ \cite{schuld2019,havlivcek2019} that are optimized with a genetic algorithm. As compared to earlier variational works\ \cite{chivilikhin2020}, we provide a comprehensive solution that automatically designs both the structure and the parameterization of the feature map circuit. Our algorithm uses an NSGA-II genetic algorithm with a multiobjective Pareto front that optimizes the accuracy and generalization power of the map, while minimizing the circuit size. We test this method against both synthetic and realistic benchmarks, finding remarkable accuracy for all problems. Moreover, the Pareto strategy and our weighting of gates seems to produce quantum feature maps that are largely uncorrelated. This hints at the possibility of constructing hybrid quantum-inspired strategies for machine learning based on these ideas.

The structure of this work is as follows. In section\ \ref{sec:qsvm} we review the method of quantum feature maps and quantum kernels for supervised learning with support vector machines. We introduce a new function\ (\ref{eq:new-kernel}) that seems to exhibit good separation properties and will be used in simulations. With this knowledge, section\ \ref{sec:GDQK} introduces the algorithm for genetically designed quantum kernels. After a brief review of genetic algorithms in section\ \ref{sec:genetic-algorithms}, we introduce an encoding of quantum feature maps as binary strings of genes. The genetic map from section\ \ref{sec:encoding} is a small example with only 5 bits, but exemplifies how to encode structure, types of gates, dependence on the input parameters and numerical parameterization of the circuit. This contrasts with other works\ \cite{chivilikhin2020} where structure and parameters were optimized separately, with different methods. In section\ \ref{sec:fitness} we describe the fitness function, which is designed for a multiobjective optimization of the accuracy, the capacity for generalization and the simplicity of the quantum feature map. We also describe how a Pareto search and elitist genetic operations can be designed to help in this optimization. Section\ \ref{sec:results} presents the results of applying our algorithm to synthetic and realistic benchmarks, in sections\ \ref{sec:toy-model} and \ref{sec:other-cases}, respectively. In these examples we also see how the optimization converges to low-entanglement feature maps, while still having good accuracy and generalization. Based on this, in section\ \ref{sec:interpretation} we discuss how such circuits could be amenable for interpretation. Finally, section\ \ref{sec:summary} summarizes the main conclusions and possible avenues for future exploration.

\section{Quantum Kernel Method}
\label{sec:qsvm}

In this work we will focus on the supervised training of binary classifiers. Given a training dataset $\{(\mathbf{x}_i,y_i)\}_{i=1}^L$ with normalized feature vectors $\mathbf{x}_i\in\mathbb{R}^d$ and binary classes $y_i\in\{+1,-1\},$ we can design a model $f(\mathbf{x})$ that predicts the class of any other point, either in this set or in unseen data. The support vector machine (SVM) is one of the earliest binary classification techniques. Developed for linearly separable data, this method constructs a hyperplane with normal $\mathbf{w}$ and displacement $b$ such that the two classes $y=+1$ and $y=-1$ lay on opposite sides of the hyperplane. The classifier has a simple form, given by a sign function
\begin{equation}
  f(\mathbf{x}) = \sign\left(\mathbf{w}^{T} \cdot \mathbf{x} + b\right).
  \label{eq:hyperplane}
\end{equation}
The hyperplane is constructed using \textit{support vectors} from the training set
\begin{equation}
  \mathbf{w} = \sum_i \beta_i y_i \mathbf{x}_i\,,
\end{equation}
in a way that maximizes the margin between those vectors and the hyperplane.

There are various techniques that turn SVM into a universal classifier, working data that is not linearly separable. One is to construct additional features or variables out the original vectors, enlarging the dimension of the classification space $\tilde{\mathbf{x}}_i := \bm{\Phi}(\mathbf{x}_i) \in \mathbf{R}^{r},$ with $r\gg d.$ By raising the dimensionality, the so called \textit{feature map} can transform the problem into a linearly separable one. Interestingly, the classifier can be inferred from a \textit{kernel function} that encodes the scalar product between the new features
\begin{equation}
  K(\mathbf{x},\mathbf{x}') = \bm{\Phi}(\mathbf{x})^T \bm{\Phi}(\mathbf{x}').
  \label{eq:kernel}
\end{equation}
This can be seen from the expression of the hyperplane $\mathbf{w}$ in terms of the new features, and how this all fits into the final classifier
\begin{eqnarray}
  \label{eq:funcion_decision}
  f(x)
  & = &
        \left(\sum_{i} \beta_i y_i \bm{\Phi}(\mathbf{x}_i)\right)^T \bm{\Phi}(\mathbf{x}) + b
        = \sum_i \beta_i y_i \bm{\Phi}(\mathbf{x}_i)^T \bm{\Phi}(\mathbf{x}) + b \\
  & = & \sum_i \beta_i y_i K(\mathbf{x}_i,\mathbf{x}) + b. \nonumber
\end{eqnarray}
Moreover, by Mercer's theorem\ \cite{geron2019hands}, we do not need to know the form of the feature map---which may even be an infinite-dimensional function---, but just a kernel function $K(\mathbf{x},\mathbf{x}')$ that has the right positivity properties to encode a scalar product.

When developing quantum classifiers for classical data, the usual approach is to engineer a feature map from classical to quantum features $\ket{\Phi(\mathbf{x})}$\ \cite{schuld2019,havlivcek2019}. This map can be trivial---e.g. encoding data into quantum register states or quantum amplitudes---, but more generally it is a parameterized unitary transformation, built from quantum gates that are depend on the input features $\ket{\Phi(\mathbf{x})} := \mathcal{U}(\mathbf{x};\bm{\theta})\ket{0}^n,$ and on some additional controls $\bm{\theta}.$ The feature map can be combined with further classification circuits or measurements, to create the so called \textit{quantum neural networks.} However, as argued in Refs.\ \cite{schuld2019,havlivcek2019,schuld2021},  we could simply use those circuits to evaluate a kernel
\begin{equation}
  K(\mathbf{x}, \mathbf{x}') =\left\vert\braket{\Phi (\mathbf{x})|\Phi(\mathbf{x}')}\right\vert^2 = \left|\braket{0^n |  \mathcal{U}(\mathbf{x};\bm{\theta})^\dagger \mathcal{U}(\mathbf{x}';\bm{\theta})|0^n}\right|^2,
\end{equation}
and derive the corresponding SVM classifier $f(\mathbf{x}; \bm{\theta}).$

In this work we use a different type of quantum kernel
\begin{equation}
  K(\mathbf{x}, \mathbf{x}') = \mathrm{Re}\braket{\Phi(\mathbf{x})|\Phi(\mathbf{x}')} = \mathrm{Re}\braket{0^n |  \mathcal{U}(\mathbf{x};\bm{\theta})^\dagger \mathcal{U}(\mathbf{x}';\bm{\theta})|0^n}.
  \label{eq:new-kernel}
\end{equation}
By not squaring the scalar product between vectors, the function resembles more the original motivation for $K(\mathbf{x},\mathbf{x}').$ Moreover, as we have confirmed numerically, this kernel allows for sharper separations and more easy convergence of the optimizer. However, while this choice is neutral from a classical simulation point of view, it is more complicated to evaluate in a quantum computer. Unlike Ref.\ \cite{havlivcek2019}, to estimate $K(\mathbf{x},\mathbf{x}')$ we would need to use an ancillary qubit, prepared in a quantum superposition, and controlling the $\mathcal{U}$ operation, to estimate the scalar product as the result of an interference process.

This quantum kernel method is usually combined with some kind of iterative update of the parameters $\bm{\theta},$ to maximize a cost function that includes the accuracy of the model and some other regularizations. Depending on the expressive power of the underlying feature map, this approach can lead to \textit{barren plateaus} and other obstacles that prevent a good training. For that reason, in this work we explore a global optimization method that trains both the parameters $\bm{\theta} $ \emph{as well as the structure} of the quantum circuit $\mathcal{U},$ by using evolutionary artificial intelligence techniques, also known as \textit{genetic algorithms.}

\section{Genetically Designed Quantum Kernels}
\label{sec:GDQK}

\begin{figure}[ht]
\centering \includegraphics[width=13cm]{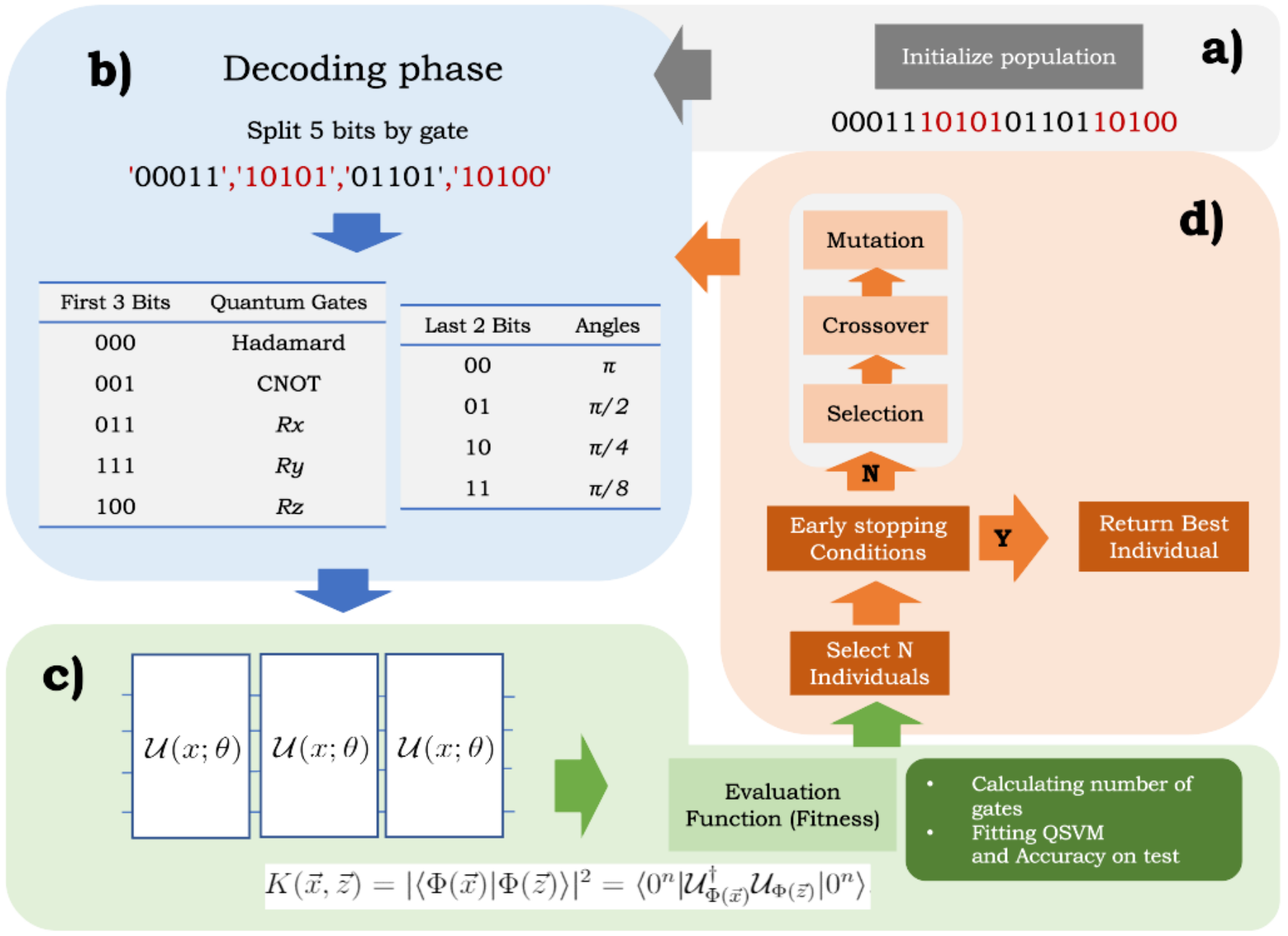}
\centering \caption{Technique scheme. \textbf{a)}. The initial population is initialized. \textbf{b)}. We can see the decoding process based on five bits per quantum gate and angle. \textbf{c)}. This individual is used in the fitness function as feature map for QSVM in order to calculate accuracy and number of gates. \textbf{d)}. The genetic algorithm process continue with phases already seen until early stopping condition so that the most optimized ansatz for this dataset is provided.}
\label{fig:coding1}
\end{figure}

\subsection{Overview of Genetic Algorithms}
\label{sec:genetic-algorithms}

Genetic algorithms are meta-heuristic optimization techniques based on the theory of evolution. The algorithms perform a guided search in a space of solutions, evolving a population of individuals with encoded the feature maps, through the application of \textit{genetic operations}. In each algorithm iteration or \textit{generation}, the resulting \textit{offspring} is selected in order to improve one or more objetives. As a result of evolutionary pressure, the collective is more likely to select the best suited individuals from a very large configuration space, in a efficient way.

A very important ingredient in the genetic algorithm is the \textit{fitness} function. This function depends on some metric that we wish to maximize (or minimize), as well as other regularizations. However, as we show in this work, it is possible for a genetic algorithm to achieve more than one goal, performing a multiobjective optimization process. In this case, the individuals in each generation that maximize (or minimize) the fitness function are called the \textit{Pareto front.} More precisely, a solution $x$ is called dominated by another solution $y$ only if $y$ is equal or better than $x$ solution with respect to all the objectives of the function. The Pareto front is the set of non-dominated solutions which satisfy all objectives defined in the fitness function, and which is progressively improved by the evolution algorithm.

A key ingredient for the success and utility of a genetic algorithm is the choice of genetic operations that evolve the population of individuals, as seen in figure\ \ref{fig:coding1}d. The \textit{selection} operator chooses a subset of the existing population to create a new generation using the crossover and mutation operators. The \textit{mutation} operator randomly alters the information of selected individuals to explore places far from the solution space. The \textit{crossover} is an stochastic operation that allows even more drastic explorations, by allowing two individuals to exchange their genetic information. Note that, while the probability of selection is proportional to the fitness of the individuals, the mutation and crossover probabilities are fixed values that have been tuned for performance.

Finally, the genetic algorithm typically involves some \textit{stopping early conditions,} statements which determine whether the evolution process has achieved its goal. Some possible strategies are checking the convergence or saturation of the fitness objetive, a minimum accuracy threshold that keeps the process going for further steps or defining a maximum number of generations.

\subsection{Genetic quantum feature map}
We will now describe a multiobjective genetic algorithm that automatically designs and optimizes quantum classifiers based on quantum feature maps and support vector machines. The algorithm explores a configuration space of parameterized quantum circuits that potentially represent feature maps. It looks for those circuits that, once trained in a QSVM method, maximize the accuracy with which they generalize to the validation data set, while minimizing the complexity of the circuit, which is measured in terms of circuit depth and difficulty of operations.

The complete algorithm is summarized in figure\ \ref{fig:coding1}. The process starts with an initial population of individuals, represented by bit strings of size $M \times N \times 5$. The evaluation function decodes each individual, creating an associated quantum circuit (see section\ \ref{sec:encoding}). This circuit, together with the training dataset, is used to implement a quantum kernel SVM algorithm, computing the fitness function (see section\ \ref{sec:fitness}). The best individuals have more probability to be elected and subjected to the different genetic operations, such as mutation, crossover and selection, creating a new generation of individuals or quantum circuits. The whole process is repeated until we meet the convergence criteria.

\subsubsection{Encoding}
\label{sec:encoding}

The first step for engineering our algorithm is to design a map from the genetic information to the quantum circuit that we wish to characterize. In our model, the genes will be binary strings that encode local, entangling and parameterized quantum gates. To create a minimalistic encoding that exemplifies all types of gates, we use five bits per gene $s_0s_1s_2s_3s_4,$ as shown in figure\ \ref{fig:coding1}b.  We aim to create a quantum circuit acting on $M$ qubits with a maximum of $N$ layers and use $M\times N\times 5$ bits in total. For simplicity, the order of actuation of the genes is sequential. Thus the $i$-th gene operates on the $j$-th qubit of the quantum register and possibly depends on the $k$-th a variable from the input data $\mathbf{x}\in\mathbb{R}^d$, with $j=i\mbox{ mod }M$ and $k=i\mbox{ mod }d.$

The mapping from bits to gates is also very straightforward. The first three bits $s_0s_1s_2$ determine whether the gate is fixed---a Hadamard or a CNOT gate---, or whether it is a local rotation parameterized by a value in the input data, $R_\alpha(\theta_i x_k)=\exp(-i\theta_ix_k\sigma^\alpha_k).$ When the gate is parameterized, the first bits $s_0s_1s_2$ select the rotation axis, the last two bits select a proportionality parameter $\theta_i=\frac{\pi}{4}2^{-2^{s_3}-4^{s_4}}.$ When the gate is a CNOT, it acts on consequitive qubits, $j$ and $j+1\mbox{ mod } M.$ All other combinations of bits not reflected in figure\ \ref{fig:coding1}b are taken to be just identity operations.

This gates selection has been chosen in such a way that all the elements found in variational circuits used as feature maps are represented, being local parameterized, non-parameterized and entanglement gates.
\subsubsection{Fitness Function}
\label{sec:fitness}

Our fitness function is designed to maximize the accuracy and minimize the complexity of the variational circuit. To measure the latter, we introduce a \textit{size metric,} labeled SM, which assigns different costs to the number of local gates $N_\text{local}$ and the number of entangling gates $N_\text{CNOT}$, weighted as follows
\begin{equation}
\text{Size Metric (SM)} = \frac{N_\text{local} + 2 N_\text{CNOT}}{N_\text{qubits}}.
\label{eq:gates}
\end{equation}
The second ingredient in the fitness function is the \textit{accuracy} of the encoded circuit. To compute this metric, we divide the data into a training set and a test set. We use the quantum circuit and the training set to compute the classifier $f(\mathbf{x})$ in the quantum kernel SVM. We then estimate the accuracy of the model $f(\mathbf{x})$ over the test set, as the fraction of points that are properly classified.

We aim to maximize both quantities in a multiobjective optimization process, creating a \textit{Pareto front} that carefully balances the relative importance of both figures of merit. A high weight in accuracy can produce a collapse into one individual loosing the necessary genetic diversity to be able to minimize the quantum circuit size along the evolution. On the other hand, a very small number of gates can hinder the power of the quantum kernel to separate the features. In order to achieve a proper balance, we engineer a fitness function that increases the relevance of the SM as the accuracy approaches its limiting value 1, using the following multiobjective fitness function
\begin{equation}
\text{Weights Control (W_C) } = \text{SM} + \text{SM}* \text{accuracy}^2.
\label{eq:weights}
\end{equation}

\subsubsection{Genetic Operators}
In the genetic algorithm we use selection, mutation and crossover operators. The selection operator is a multiobjetive \textit{Non-Sorted Genetic Algorithm II} (NSGA-II). This algorithm decides which individuals survive to the next generation based on Pareto-dominance and density-based metrics \cite{baran2021}. This algorithm has a strong tendency to keep individuals with higher fitness, because it selects the individuals after ordering the population by dominance. However, it also uses the non-dominated individuals' rank and density-distance to add diversity in each generation.

\begin{figure}[t]
\noindent\raisebox{-5cm}{\includegraphics[width=0.85\linewidth]{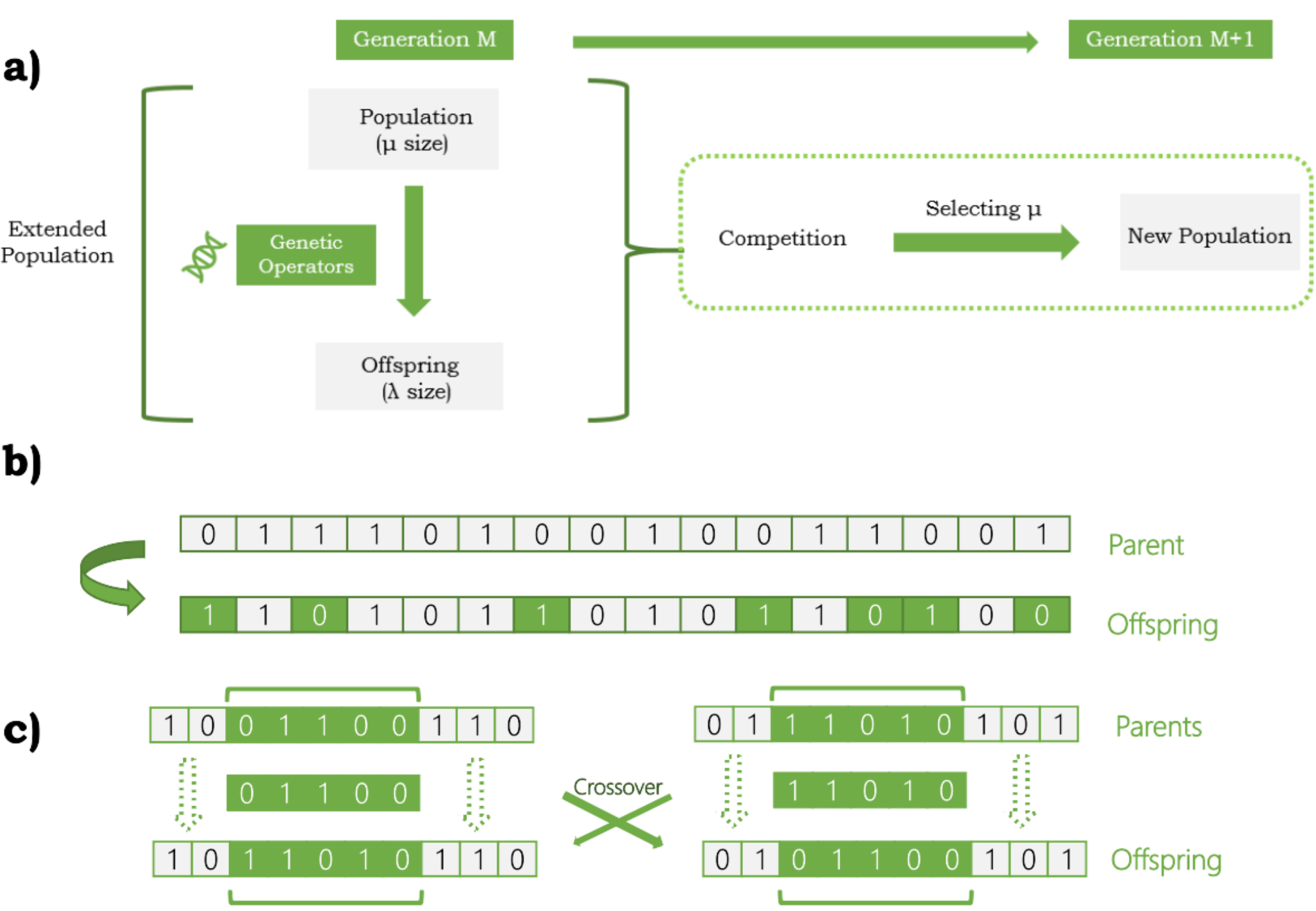}}\hfill\\
\centering \caption{Genetic operators. (a) Genetic algorithm $\mu + \lambda$ strategy. The initial population $\mu$, after applying the genetic operators, produces an offspring of $\lambda$ size. This offspring competes with the parents and from this competition a number of individuals equal to the size of the individual population $\mu$ is selected. This process is repeated throughout the generations of evolution. (b) Mutation. A \textit{bit flip} mutation affects a fraction of the genes , giving rise to a different individual. (c) Crossover. Randomly selected genes (marked in green) are exchanged between two individuals, keeping the rest of the genes chain constant.}
\label{fig:genetic-operators}
\end{figure}

Since our feature maps are coded in binary format, we can use a \textit{bit flip} mutation operator (c.f. figure\ \ref{fig:genetic-operators}b). The value $p_\text{mut}$ indicates the probability of an individual to be mutated. Once an individual is selected for mutation, each gene can be flipped with a probability $p_\text{ind}.$ As for the crossover operator, we implement a binary swap of contiguous bit substrings (see figure\ \ref{fig:genetic-operators}c). The value $p_\text{cross}$ determines probability that a crossover takes place, while the beginning and end of the swapped bits are randomly chosen along the complete strings.

We increase the elitism of the algorithm with a Mu Plus Lambda ($\mu + \lambda$) algorithm. This strategy modifies how we create the next generation of individuals, establishing a competition between the current population (size $\mu$) and the offspring ($\lambda$) that is obtained by genetic operators, as sketched in figure\ \ref{fig:genetic-operators}a. This competition ensures genetic diversity while it also preserves the best individuals that have been obtained through the evolutionary algorithm\ \cite{geneti_algo_pyth}.

All hyperparameters previously mentioned ---crossover and mutation probabilities and population sizes---were optimized and tested, to achieve a good compromise between convergence speed and optimal classification. After several tests, the optimal hyperparameters were found to be 30\% probability of crossover and 70\% mutation, with a 20\% probability of bits being mutated. This is an interesting balance that allows exploring drastic changes in the population through crossover, while maintaining a high rate of small changes through mutation. While this may seem very aggressive, the random component is kept in check by the high elitism of the competition between children and parents in the Mu Plus Lambda strategy.

\section{Results and Discussion}
\label{sec:results}

\subsection{Toy model}
\label{sec:toy-model}

\begin{figure}[t]
  \begin{flushleft}
 \textbf{a)} \raisebox{-2.5cm}{\begin{minipage}{0.5\linewidth}
    \includegraphics[width=\linewidth]{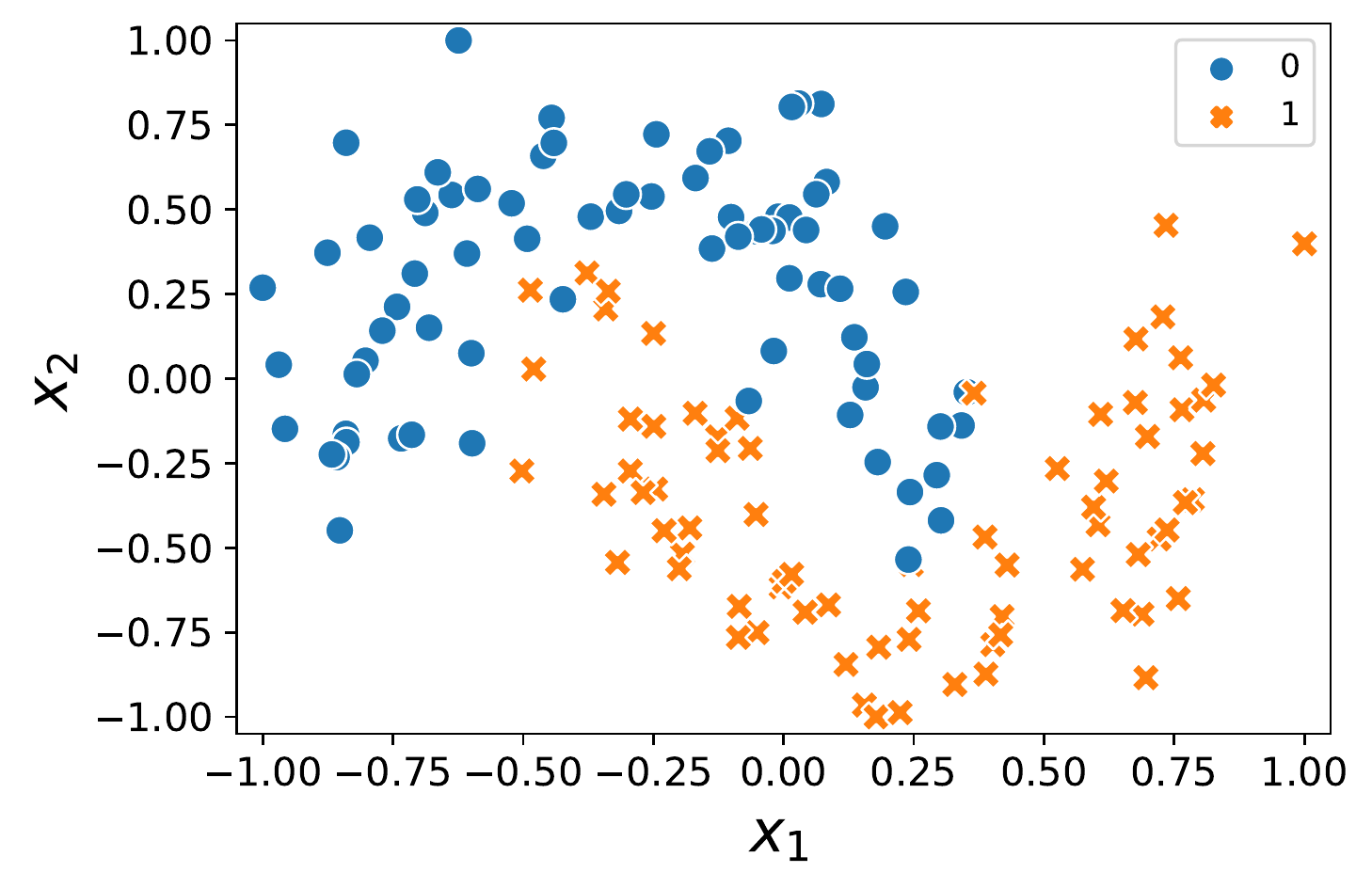}
  \end{minipage}}~~%
\textbf{b)}\raisebox{-2.5cm}{
  \begin{minipage}{0.3\linewidth}
  \begin{tabular}{lc}
               \hline
Variables&  Value \\\hline \hline
Generations & 5000 \\\hline
Population ($\mu$) & 100 \\\hline
Offspring ($\lambda$)& 15 \\\hline
Qubits     & 6 \\\hline
Max. Layers& 6 \\\hline
Crossover Prob. & 0.3 \\\hline
Mutation Prob.  &  0.7 \\\hline
Mutation Prob. Ind.  &  0.2 \\\hline
  \end{tabular}
\end{minipage}
}\\
\noindent
\textbf{c)}~~ \raisebox{-4cm}{\includegraphics[width=0.4\linewidth]{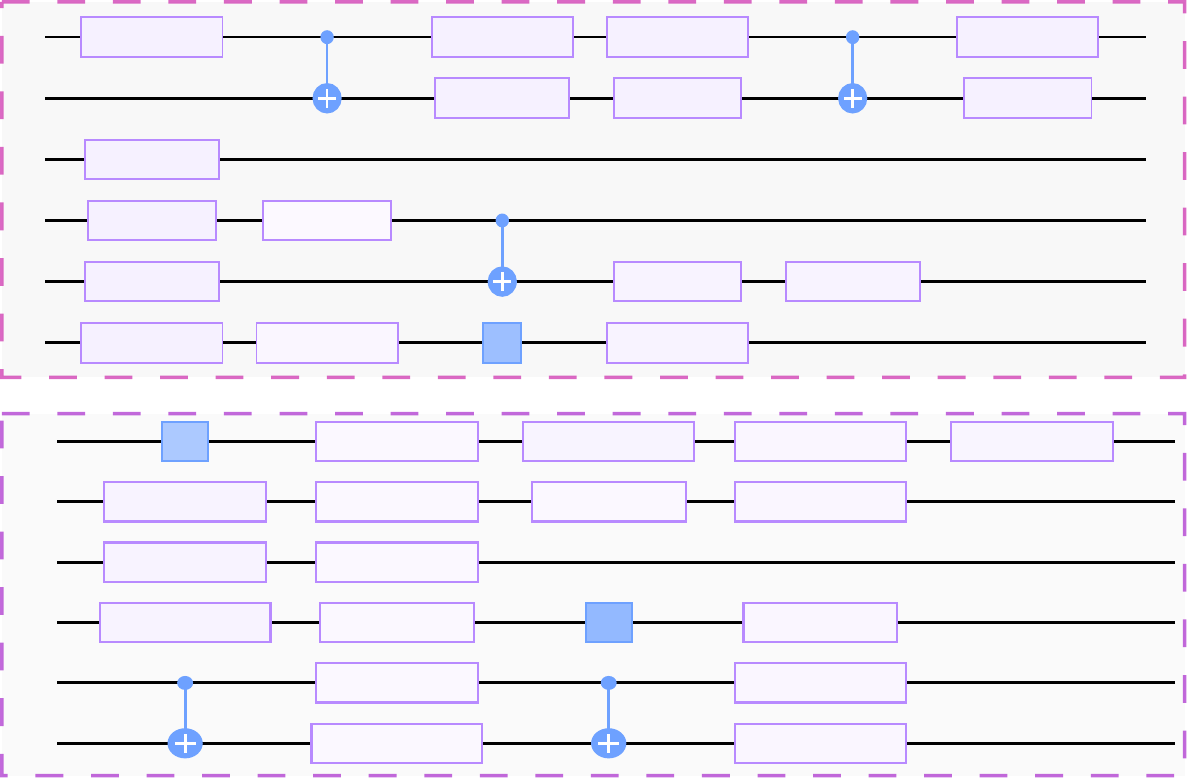}}\quad\quad
\textbf{d)}
\raisebox{-3cm}{\includegraphics[width=0.35\linewidth]{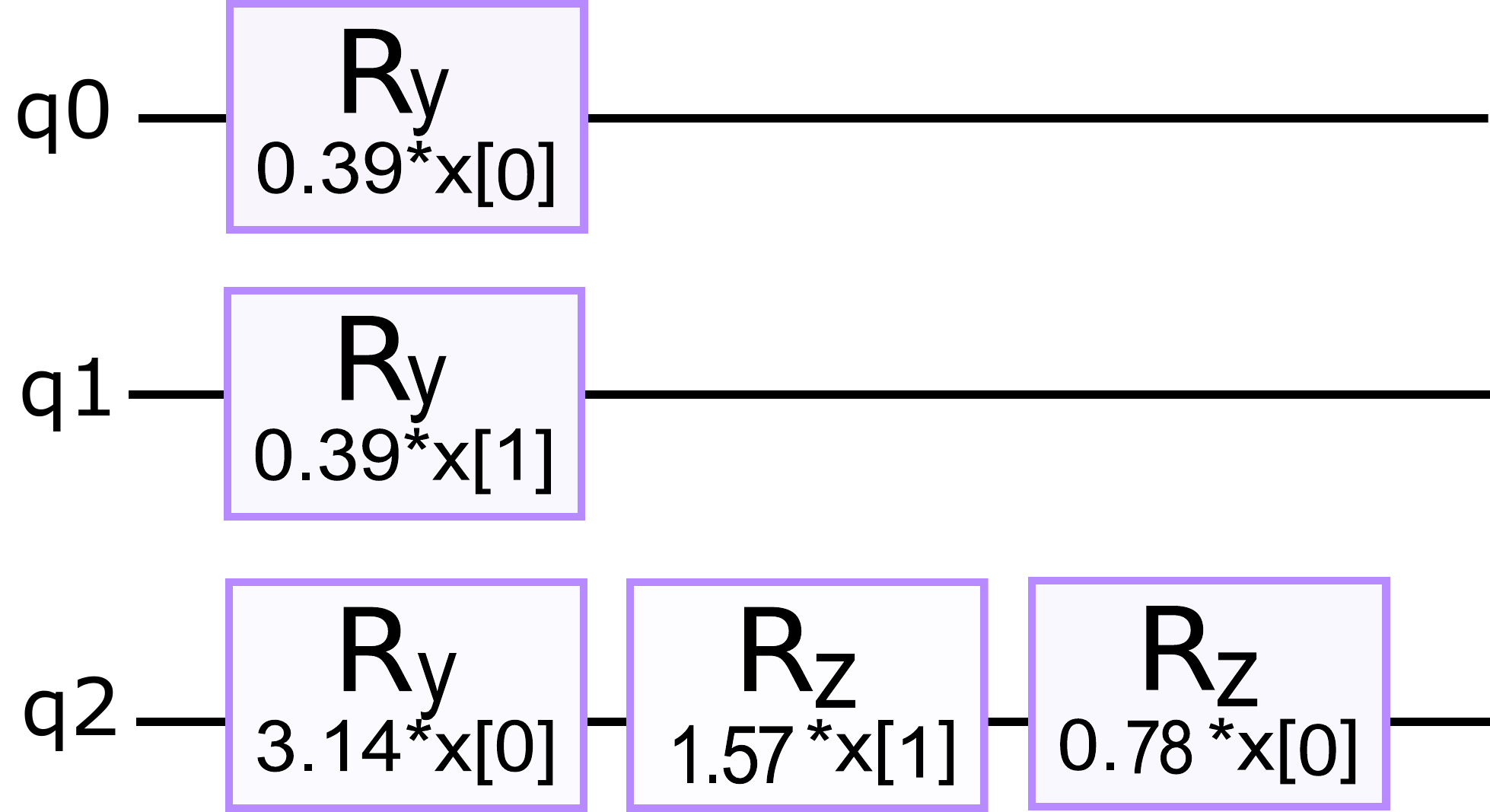}}
\end{flushleft}
\centering \caption{(a) Dataset composed by 150 points with a non-linear pattern and a binary target. (b) Hyperparameters used to optimize the QSVM circuit with our genetic algorithm. (c) Structure of circuits created in the first generations of the genetic algorithm. (d) Final circuit with 1.0 accuracy.}
\label{fig:no_lin_ds}
\end{figure}

In order to proof this new method, we use the Moons synthetic non-linear dataset with two classes, shown in figure\ \ref{fig:no_lin_ds}a, and generated using Scikit\ \cite{scikit-datasets}.  The 150 datapoints are scaled between [-1,+1] as a preprocessing step, and randomly split into a training (70\%) and test (30\%) sets. These sets are used to train quantum circuits with up to 6 qubits and 6 layers, which are progressively optimized by the genetic algorithm. As illustrated in figure\ \ref{fig:no_lin_ds}b, we optimize the circuits over 5000 generations, using a population of 100 individuals.

The initial circuits make use of all available qubits and all layers, as shown in figure\ \ref{fig:no_lin_ds}c. Already in these circuits we observe the penalty associated to CNOT gates decreasing the number of entangling unitaries, as compared to other ansätze in the literature. More interestingly, the \textit{Pareto front} combined with the elitist strategies is capable of further realizing that no entanglement at all is required to fit this model. Thus, after 5000 generations, the algorithm produces the simple uncorrelated circuit from figure\ \ref{fig:no_lin_ds}d, which fits the test set with perfect accuracy.

\begin{figure}[t]
  \begin{flushleft}
    \textbf{a)} \raisebox{-4cm}{\includegraphics[width=0.45\linewidth]{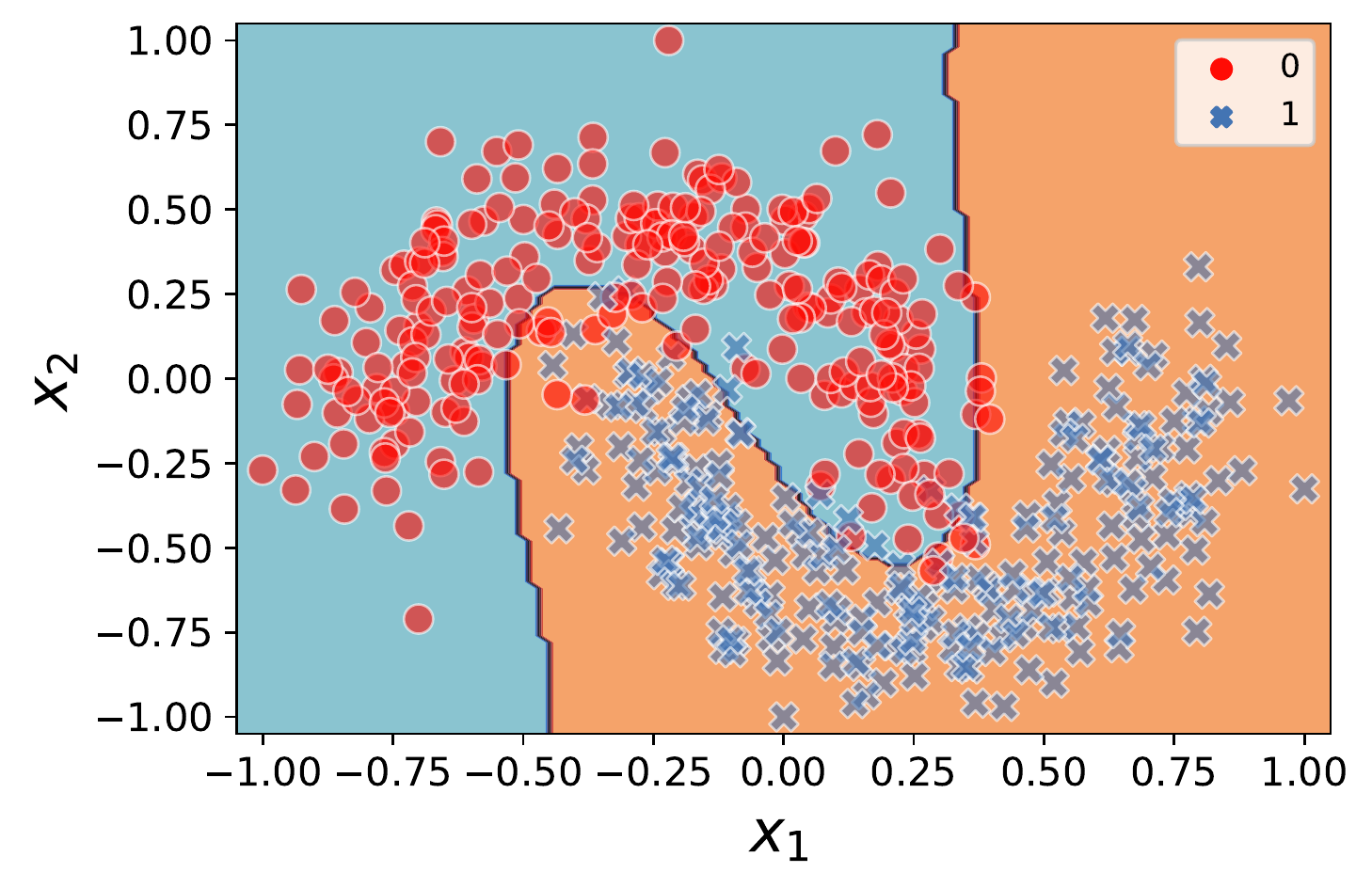}}
    \quad
    \textbf{b)} \raisebox{-4cm}{\includegraphics[width=0.3\linewidth]{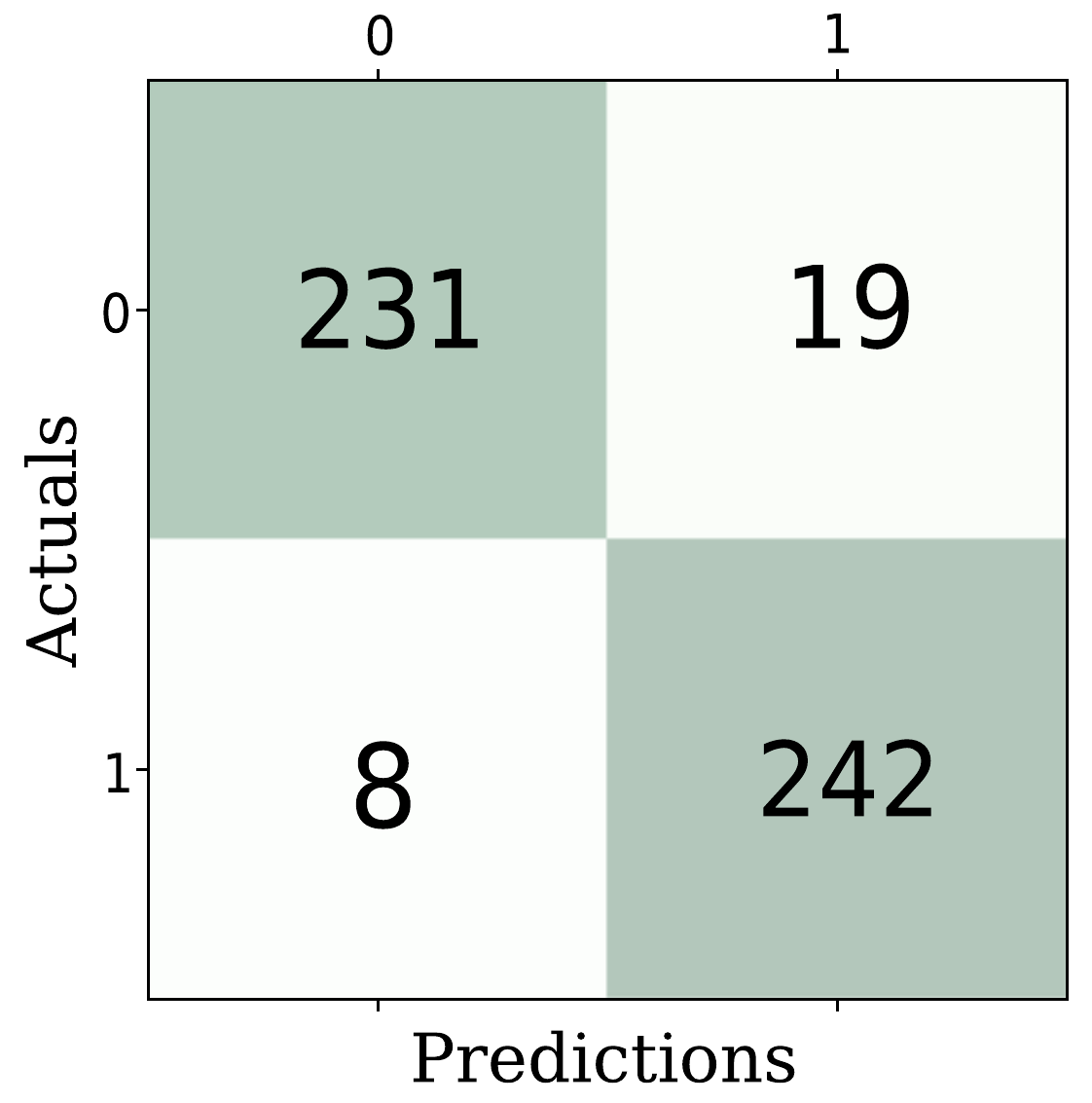}}
\end{flushleft}
\caption{(a) Validation dataset, together with the predictions and decision boundary from the generated model. (b) Confusion matrix produced by the application of the QSVM model onto the validation dataset.}
\label{fig:val}
\end{figure}

The fact that the generated model has perfect accuracy is useless, if it cannot generalize to other data from the same distribution. We \textit{validate} the utility of the model using additional datapoints, a \textit{validation set,} with 500 points generated by the same synthetic algorithm. The same scaling preprocessing step [-1,+1] that is applied to the training data is also applied to these validation datapoints. Figure\ \ref{fig:val}a shows both the validation dataset used and the predictions made by the quantum support vectorial, defined by the decision boundary. Figure\ \ref{fig:val}b also illustrates the confusion matrix of this validation process, considering both real and predicted labels, and identifying the incorrectly classified data. The confusion matrix allows us to conclude that the QSVM extrapolates to unseen data with same distribution, because 473 out of 500 data in the dataset have been correctly classified. In other words, a 94.6\% of correct classified data or 0.946 accuracy.

\subsection{Interpretability}
\label{sec:interpretation}

\begin{figure}[t]
  \begin{flushleft}
    \raisebox{-4cm}{\includegraphics[width=0.46\textwidth]{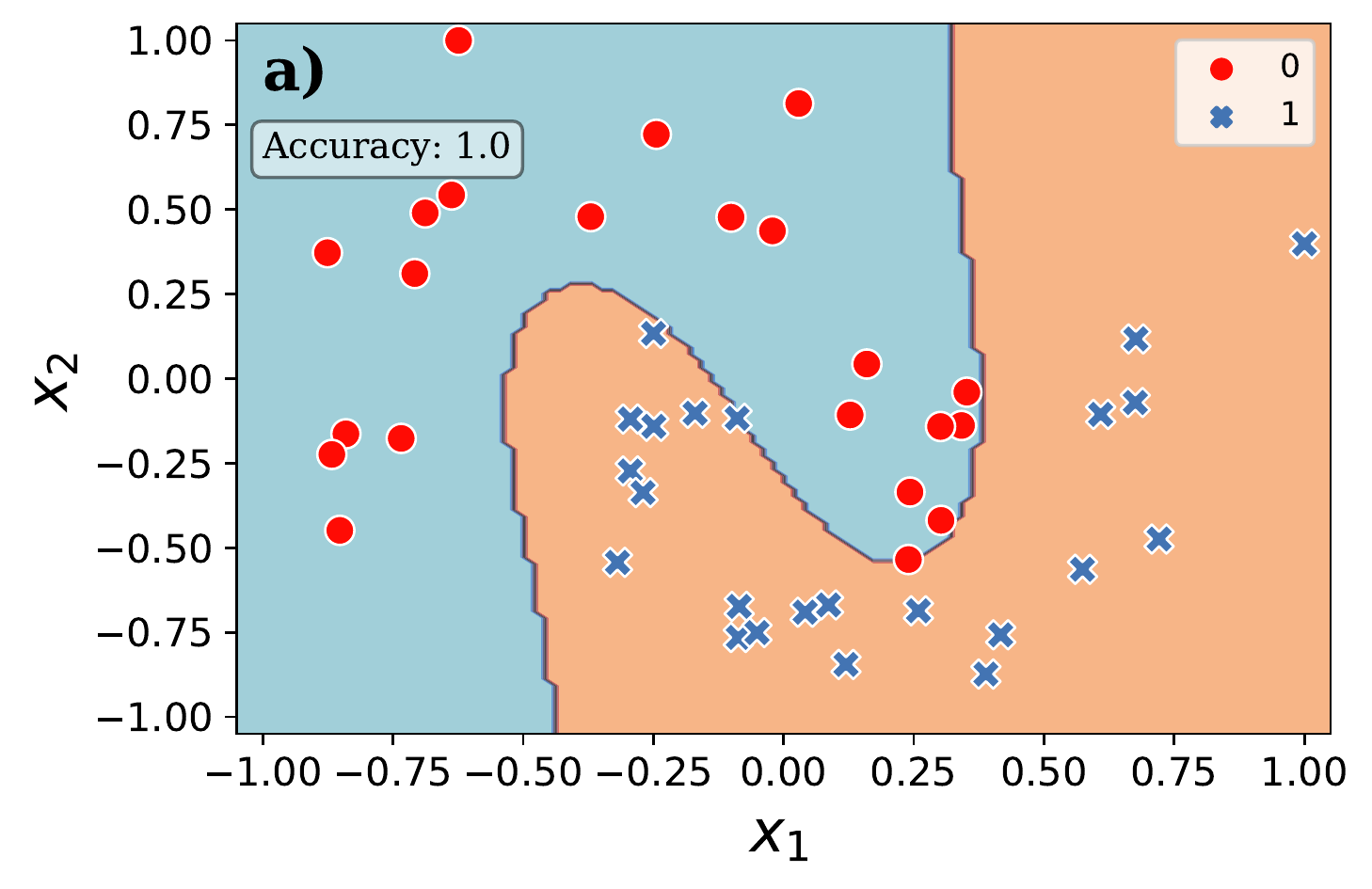}}
    \raisebox{-4cm}{\includegraphics[width=0.45\textwidth]{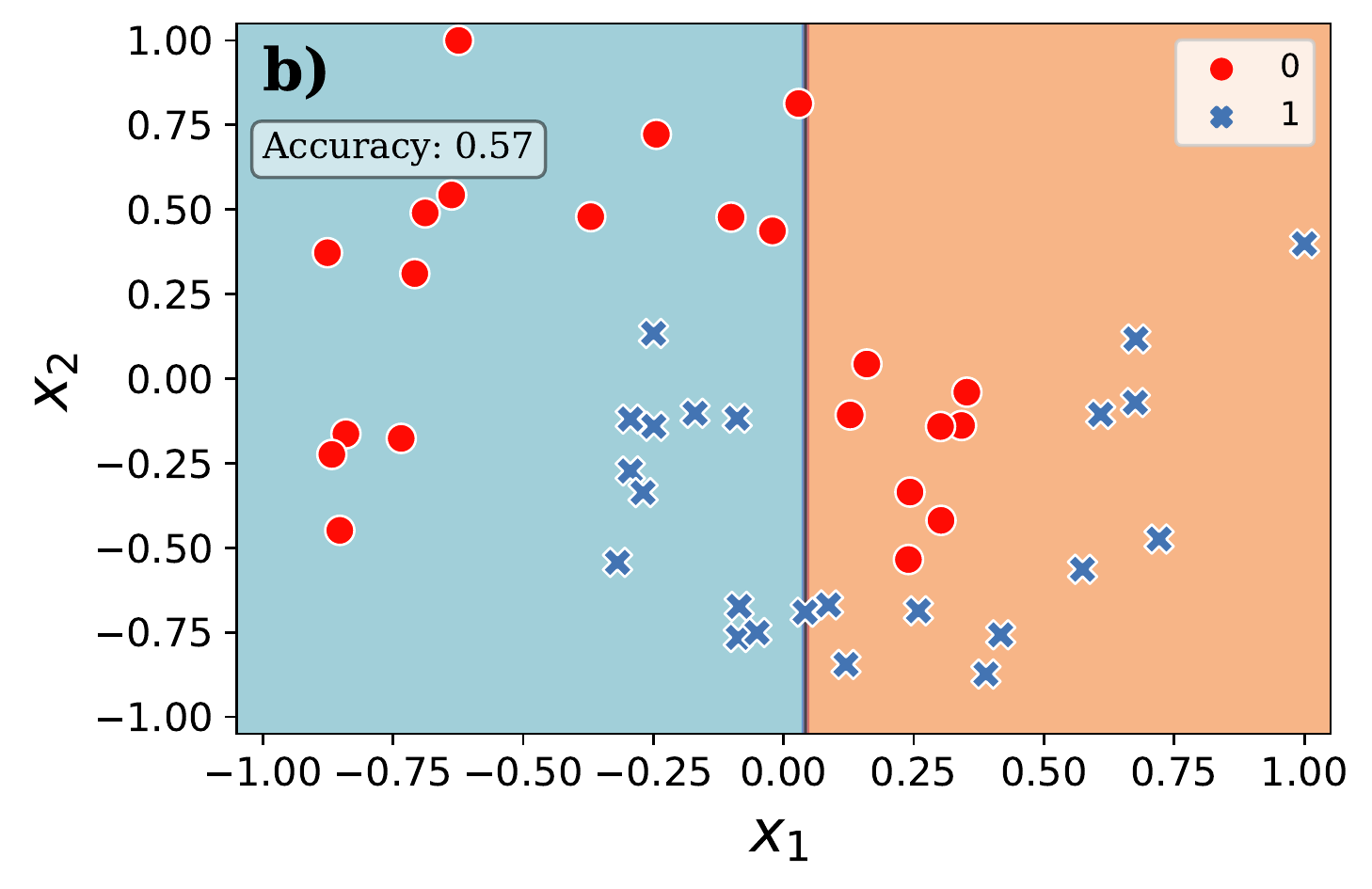}}\\
\raisebox{-4cm}{\includegraphics[width=0.46\textwidth]{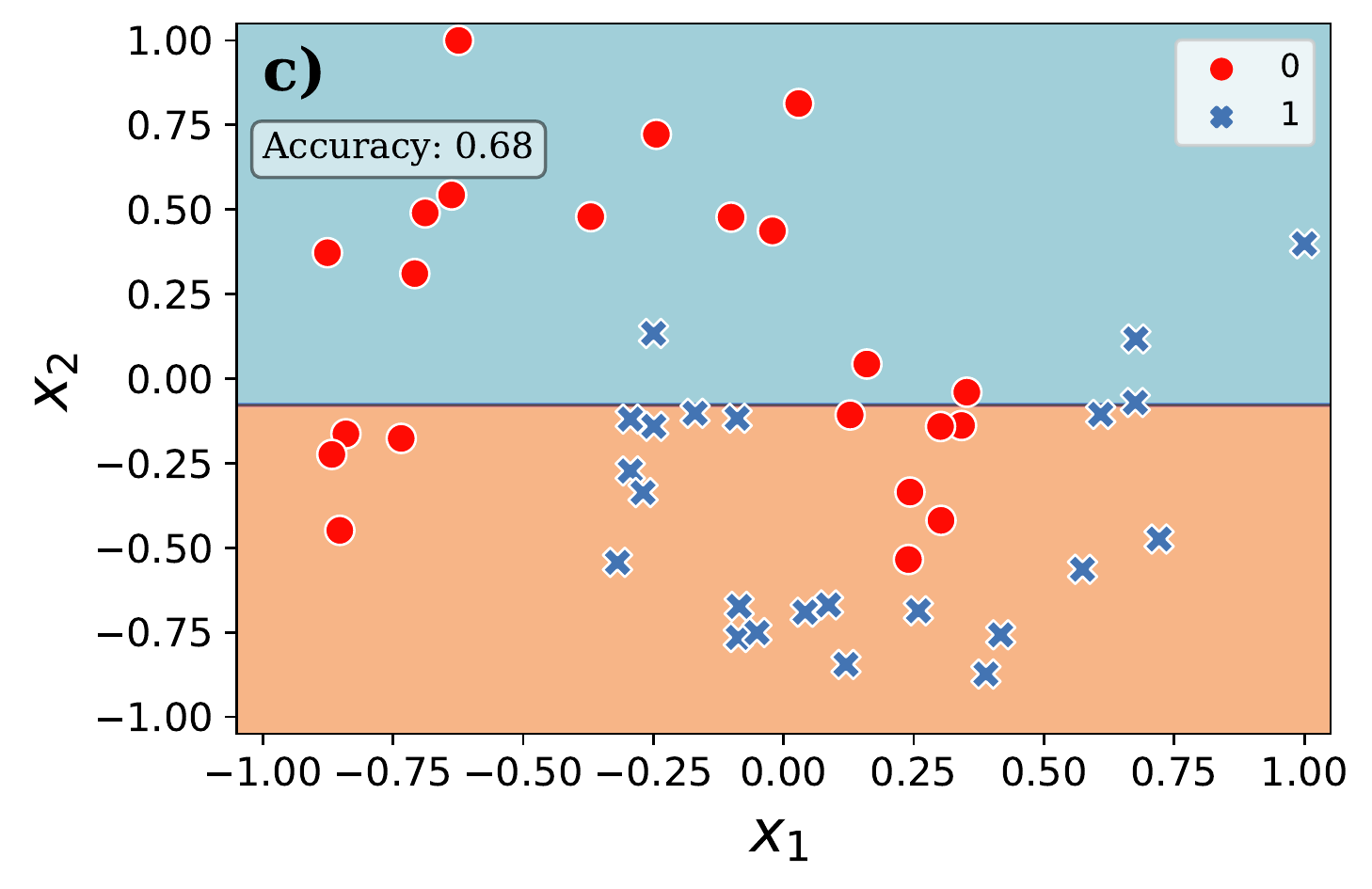}}
    \raisebox{-4cm}{\includegraphics[width=0.45\textwidth]{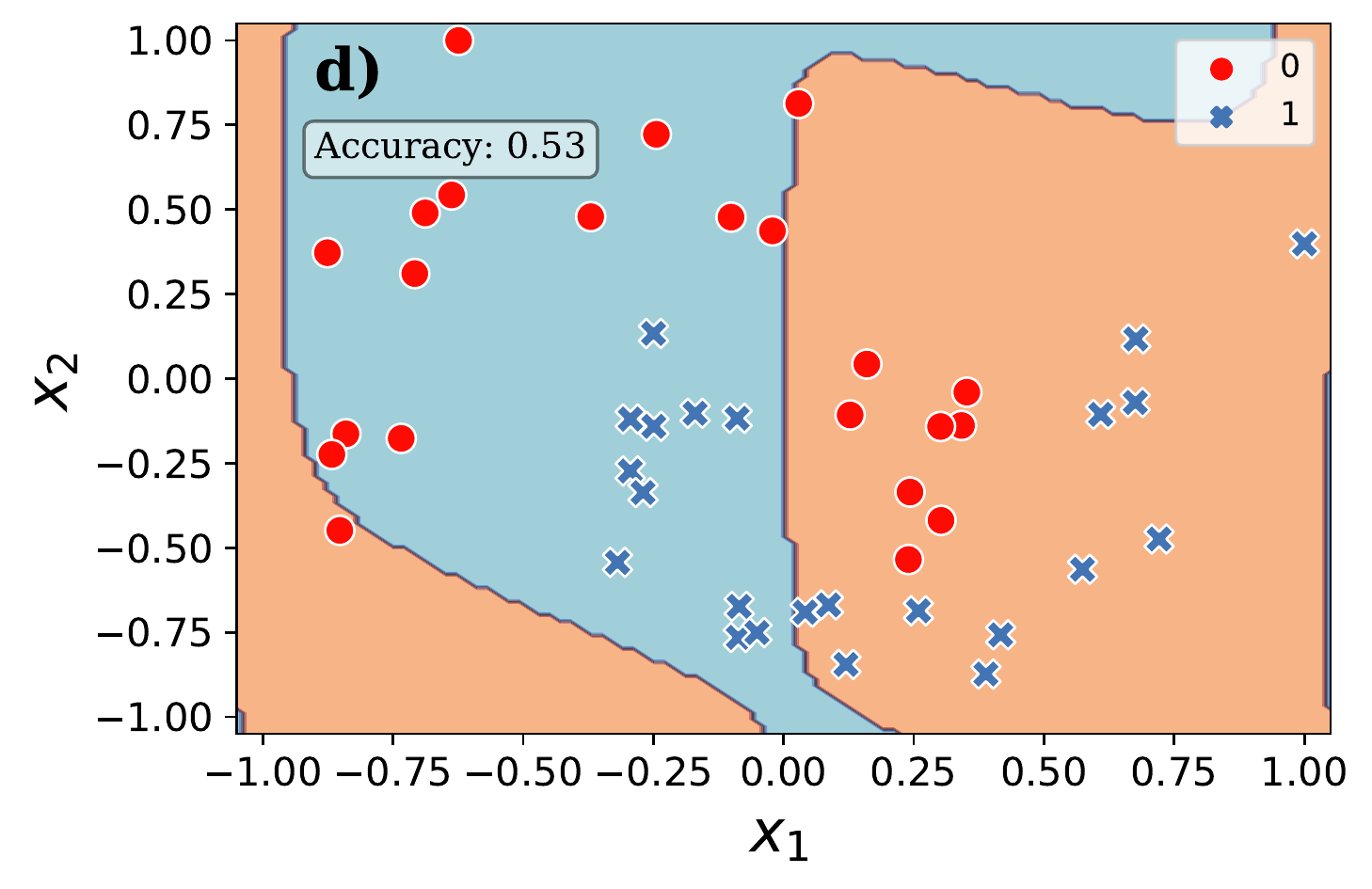}}
  \end{flushleft}
\caption{ (a) Data points and prediction boundaries from the full quantum kernel SVM. (b), (c) and (d) are the decision boundaries provided by the circuits on the first, second and third qubit, respectively.}
\label{fig:db}
\end{figure}

As we see above and will see in later examples, the strong penalty on entangling gates makes the genetic algorithm prefer circuits that have smaller clusters of uncorrelated qubits. Ideally, only the gates that are essential for the modelization are included. The result is a circuit that can be decomposed as a tensor product of separate unitaries, and a quantum kernel that is a scalar product of separate kernels, as in $K(\mathbf{x},\mathbf{x}')=\prod_{i=1}^m K_i(\mathbf{x},\mathbf{x}').$ We suggest to study the classification induced by each kernel $K_i$ separately and by their combination, as a strategy to provide \textit{interpretations} of the rules that the evolutionary strategy has produced.

An example of this study is performed in figure\ \ref{fig:db} for our synthetic model. Figure\ \ref{fig:db}a illustrates the boundaries of the complete kernel, which has accuracy 1.0, while later figures\ \ref{fig:db}b-d show the boundaries induced by each separate kernel. As we can see, both qubits one and two, provide linear hyperplanes, being qubit three the one that provides a degree of non-linearity achieved by applying three rotations. Finally, the combination of these qubits forms the desired non-linear pattern. Interestingly, the single-qubit boundaries have a lower classification accuracy, of 0.57, 0.68 and 0.53, respectively, but their nonlinear combination in the final kernel gives the right predictions.

\subsection{Other Use cases}
\label{sec:other-cases}

\begin{table}[ht]
  \caption{Results from applying the genetic engineering of QSVM to other model problems in supervised machine learning. For the most difficult model we also provide the accuracy of other classical methods for supervised machine learning: a k-NN, a linear support vector machine, and an SVM with a polynomial kernel with degree 2 (poly).}
  \begin{tabular}{lccc}
    \hline
    & Parkinson\ \cite{park_} & IoT irrigation\ \cite{iot_cl} & Drug classification\ \cite{drug_cl} \\
    \hline
    Circuit & Fig.\ \ref{fig:other-circuits}a & Fig.\ \ref{fig:other-circuits}c & Fig.\ \ref{fig:other-circuits}b \\
    Accuracy & \textbf{1.0} & \textbf{1.0} & \textbf{1.0} \\
    Generations & 5000 & 1000 & 500 \\
    \# attributes & 22 & 2 & 5 \\
    \# classes & 2 & 2 & 5 \\
    Max qubits & 15 & 5 & 5 \\
    Max depth & 8 & 5 & 5\\
    Mutation probability ($p_\text{mut}$) & 0.7 & 0.7 & 0.7 \\
    Mutation ind. prob. ($p_\text{ind}$) & 0.2 & 0.2 & 0.2 \\
    Crossover prob. ($p_\text{cross}$) & 0.3 & 0.3 & 0.3 \\
    \hline
    k-NN accuracy & 0.82 &1.0 & 0.70 \\
    SVM (linear) accuracy & 0.89&1.0 & 0.87 \\
    SVM (poly - 2) accuracy &0.89 &1.0 & 0.65 \\
    \hline
  \end{tabular}
  \label{tbl:otras_apps}
\end{table}

\begin{figure}[t]
\textbf{a)}~\raisebox{-6.2cm}{\includegraphics[width=0.9\linewidth]{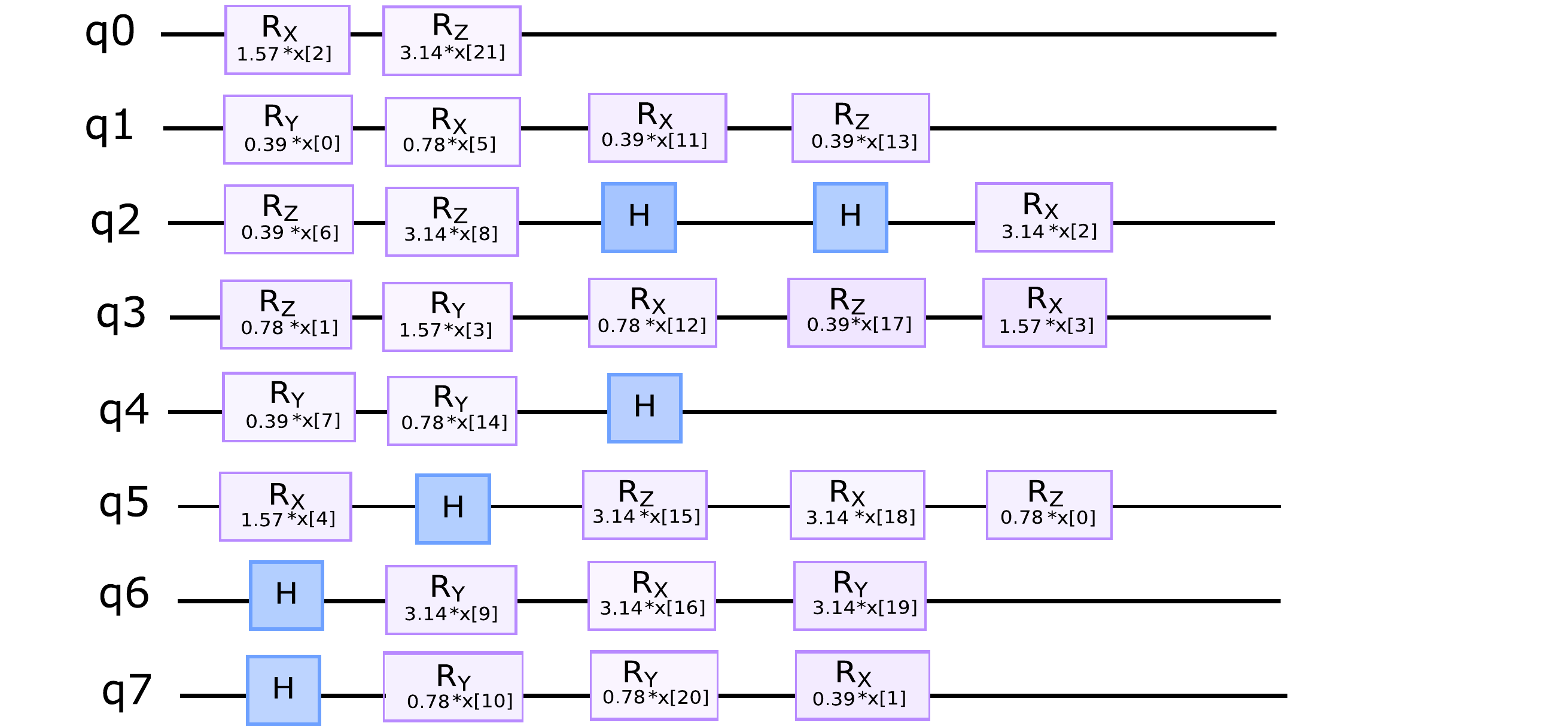}}\\[0.5cm] \textbf{b)}~\raisebox{-4cm}{\includegraphics[width=0.8\linewidth]{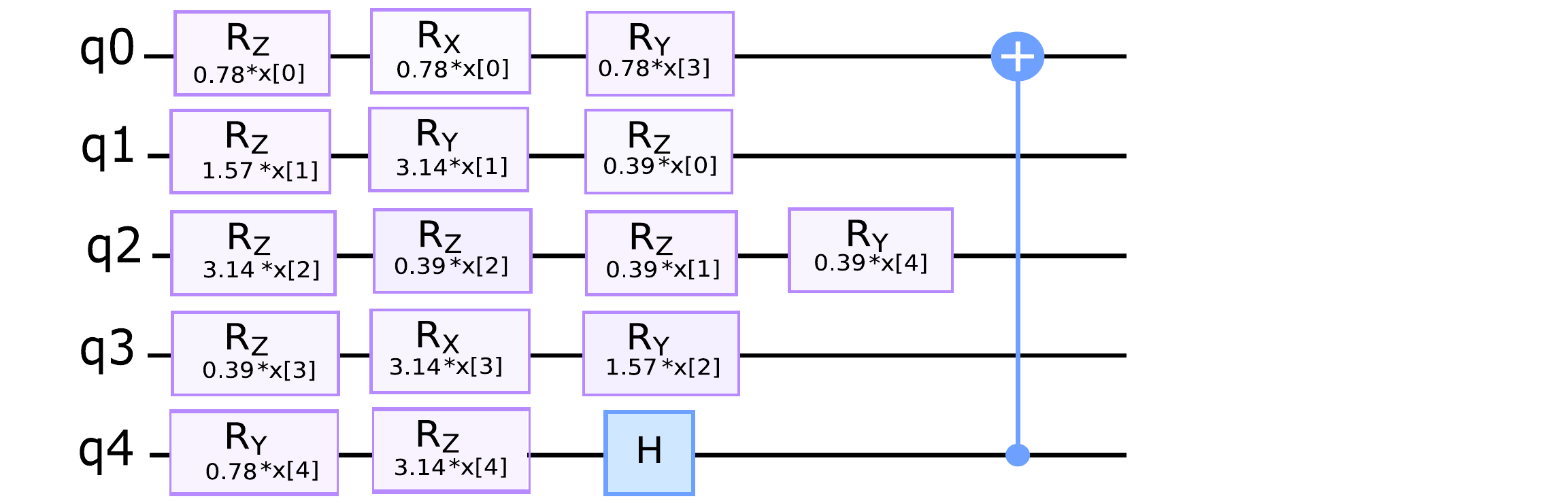}}\\[0.5cm]
\textbf{c)}~\raisebox{-2cm}{\includegraphics[width=0.3\linewidth]{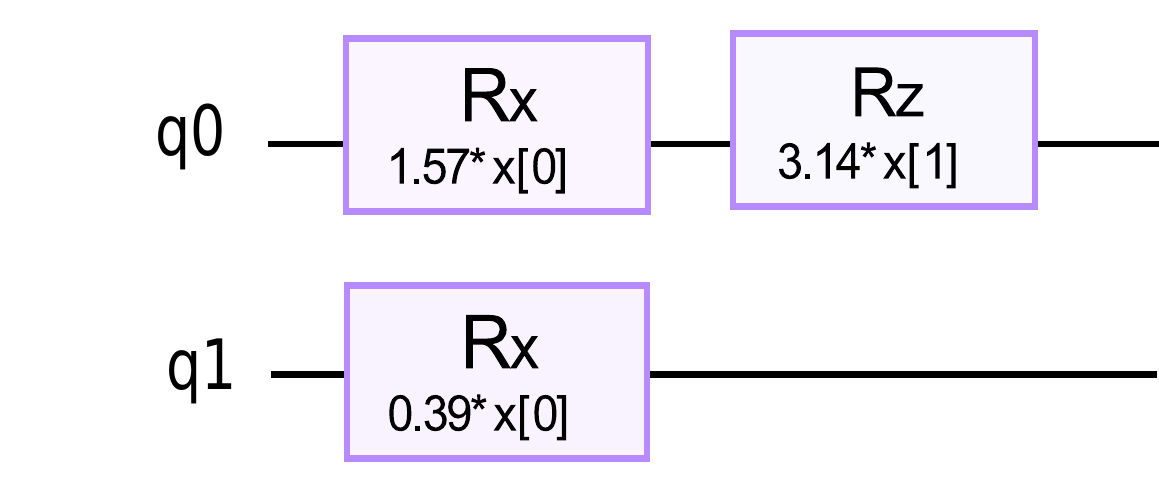}}
  \caption{Circuits generated for the supervised learning problems in Table\ \ref{tbl:otras_apps}. (a) Parkinson problem\ \cite{park_}, (b) drug classification\ \cite{drug_cl} and (c) IoT irrigation\ \cite{iot_cl}.}
  \label{fig:other-circuits}
\end{figure}

We have applied the method to other problems that are standard benchmarks for classical supervised learning techniques. Table\ \ref{tbl:otras_apps} lists three problems, with the characteristics of the datasets, the hyperparameters of the genetic algorithm and the resulting accuracy. As seen in our experiments, the technique performs well for datasets with a high number of features as well as for datasets with more than two classes. The comparison with non-quantum classification methods also shows and advantage of the QSVM technique, specially in the highly complex multiclass classification of drugs.

Figure\ \ref{fig:other-circuits} illustrates the structure and parameterization of the quantum feature maps that optimally classify these benchmarks. Interestingly, two of the circuits are uncorrelated and have no CNOT gates, while the third one, for multiclass drug classification, has just one entangler gate. This is relevant for several reasons. First, it illustrates the power of individual qubits a quantum classifiers, a realization already introduced in Ref.\ \cite{perezsalinas2021}. Second, the structures we have obtained, having little or no correlation, admit an efficient classical simulation which constitutes in itself a type of \textit{quantum-inspired} machine-learning technique.

\section{Summary and Outlook}
\label{sec:summary}

In this work we have explored the global optimization of quantum feature maps in a quantum kernel SVM algorithm using evolutionary multiobjective algorithms. The feature map is built as a parameterized quantum circuit that depends on the input data. The genetic algorithm stored the structure of the circuit, the actual gates and the functional dependence on the data as a string of binary-encoded genes. The algorithm evolves a population of individuals with genetic operators that seek to maximize the accuracy of these feature maps in modeling the data, while minimizing the complexity of the circuit. This is implemented using a nonlinear fitness function that combines both goals, and simultaneously applying a Pareto front selection strategy for the individuals.

We have applied this algorithm both to synthetic and to realistic benchmarks in the field of supervised machine learning, both single- and multiclass classification. The algorithm produces 100\% accurate classifiers that can still generalize to unseen data since this metric is obtained from test sets. Moreover, the classifiers have a simple structure, with minimal or no correlations, which still capture the underlying nonlinear patterns. We attribute the simplicity of these circuits to the classification power of single qubits and single-qubit operations\ \cite{perezsalinas2021}, which is enhanced by the combination of multiple parallel circuits. We believe that the resulting circuits are amenable to further interpretation strategies, in a simpler way than neural networks or other ML ansätze. Moreover, our results suggest the power of product states as another quantum-inspired variational strategy for supervised learning.

Our work leaves many avenues for exploration. The gene encoding that we have implemented contains a minimalistic set of entangling, local and parameterized gates, with sufficient precision for the problems we have explored. This can be extended in various ways, such as enlarging the set of weights in the parameterization, changing the order in which parameters appear in the circuit, including also more local and entangling gates, including free parameters $\theta_i$ that can be optimized using SPSA or other strategies, etc. If we focus on entanglement-free ansätze, we also find a rich avenue to explore the implementation of these models as standalone tools for machine learning, or developing a more clear strategy for the interpretation of the resulting classifiers---e.g. developing a kind of rule-based explanation of the model.
\ack
The authors gratefully acknowledges the computer resources at Artemisa, funded by the European Union ERDF and Comunitat Valenciana as well as the technical support provided by the Instituto de Física Corpuscular, IFIC (CSIC-UV).

This work has been supported by Spanish project PGC2018-094792-B-100 (MCIU/AEI/FEDER, EU), CAM/FEDER Project No. S2018/TCS-4342 (QUITEMAD-CM), and CSIC Platform PTI-001.

\section{References}
\providecommand{\newblock}{}

\bibliographystyle{iopart-num}
\end{document}